\documentclass[conference]{IEEEtran}
 \IEEEoverridecommandlockouts

\usepackage{ifpdf}
\usepackage{cite}
\usepackage{color}
\ifCLASSINFOpdf
 \usepackage[pdftex]{graphicx}
\else
   \usepackage[dvips]{graphicx}
   \DeclareGraphicsExtensions{.eps}
\fi
\usepackage[cmex10]{amsmath}
\usepackage{algorithmic}
\usepackage{array}
\usepackage{mdwmath}
\usepackage{mdwtab}
\usepackage{eqparbox}
\usepackage{fixltx2e}
\usepackage{stfloats}
\usepackage{url}
\usepackage{epstopdf}
\usepackage{enumerate}
\usepackage{amsmath, amsthm, amssymb}

\newtheorem{definition}{Definition}
\newtheorem{theorem}{Theorem}

\newtheorem{claim}{Claim}
\newtheorem{remark}{Remark}

\begin{document}

\title{On the Worst-case Communication Overhead for Distributed Data Shuffling\vspace{-28pt}}

\author{Mohamed Adel Attia \hspace{25pt} Ravi Tandon\\
  Department of Electrical and Computer Engineering\\
 University of Arizona, Tucson, AZ, 85721\\
Email: \emph{\{madel, tandonr\}}@email.arizona.edu}

\maketitle

\begin{abstract}
Distributed learning platforms for processing  large scale data-sets are becoming increasingly prevalent. In typical distributed implementations, a centralized master node  breaks the data-set into smaller batches for parallel processing across distributed workers to achieve speed-up and efficiency. Several computational tasks are of sequential nature, and involve multiple passes over the data. At each iteration over the data, it is common practice to randomly re-shuffle the data at the master node, assigning different batches for each worker to process. This random re-shuffling operation comes at the cost of extra communication overhead, since at each shuffle, new data points need to be delivered to the distributed workers.

In this paper, we focus on characterizing the information theoretically optimal communication overhead for the distributed data shuffling problem. We propose a novel coded data delivery scheme for the case of no excess storage, where every worker can only store the assigned data batches under processing. Our scheme exploits a new type of coding opportunity and is applicable to any arbitrary shuffle, and for any number of workers. We also present information theoretic lower bounds on the minimum communication overhead  for data shuffling, and show that the proposed scheme matches this lower bound for the worst-case communication overhead.
\end{abstract}

%

\section{Introduction}
\label{sec:Introduction}
Processing of large scale data-sets over a large number of distributed servers is becoming increasingly prevalent. The parallel nature of distributed computational platforms such as Apache Spark\cite{ZaChFrShSt2010}, Apache Hadoop \cite{ShKuRaCh2010}, and MapReduce \cite{DeGh2004} enables the processing of data-intensive tasks common in machine learning and empirical risk analysis. In typical distributed systems, a centralized node which has the entire data-set assigns different parts of the data to distributed workers for iterative processing.

Several practical computational tasks are inherently sequential in nature, in which the next iteration (or pass over the data) is dependent on the previous iteration.
Of particular relevance are sequential optimization algorithms such as incremental gradient descent, stochastic gradient descent, and random reshuffling.  The convergence of such iterative algorithms depends on the order in which the data-points are processed, which in turn depends on the skewness of the data. However, the \textit{preferred ordering} of data points is unknown apriori and application dependent. One commonly employed practice is to perform \textit{random reshuffling}, which involves multiple passes over the whole data set with different orderings at each iteration. Random reshuffling has recently been shown to have better convergence rates than stochastic gradient descent \cite{RR-arxiv-2015,IoSz:2015}. 

Implementing random reshuffling in a distributed setting comes at the cost of an extra communication overhead, since at each iteration random data assignment is done for the distributed workers, and these data points need to be communicated to the distributed workers. This leads to a fundamental trade-off between the communication overhead, and storage at each worker. On one extreme case when each worker can store the whole data-set, no communication is necessary for any shuffle. On the other extreme, when the workers are just able to store the batches under processing, which is refereed to as the \textit{no-excess storage} case, the communication overhead is expected to be maximum.

\vspace{5pt}

\noindent \textit{\textbf{Main Contributions:}} The main focus of this work is characterizing the information theoretic optimal communication overhead for the \textit{no-excess storage} case. 
The main contributions of this paper are summarized as follows:

$\bullet$ We present an information theoretic formulation of the problem, and develop a novel approach of describing the communication problem through a shuffling matrix which describes the data-flow across the workers.

$\bullet$ We next present a novel coded-shuffling scheme which exploits a new type of coding opportunity in order to reduce the communication overhead, in contrast to existing approaches. Our scheme is applicable to any arbitrary shuffle, and for any number of distributed workers. 

 $\bullet$ We present information theoretic lower bounds on the  communication overhead as a function of the shuffle matrix. Moreover, we show that the proposed scheme matches this lower bound on the worst-case communication overhead, thus characterizing the information theoretically optimal worst-case communication necessary for data shuffling.

\vspace{5pt}
\noindent \textit{\textbf{Related work:}} The benefits of coding to reduce communication overhead of shuffling were recently investigated in \cite{Kannan-2015}, which proposes a probabilistic coding scheme. However, \cite{Kannan-2015} focuses on  using the excess storage at the workers to increase the coding opportunities and reduce the average communication overhead. In our recent work \cite{AtRa:GC2016}, we presented the optimal worst-case communication overhead for any value of storage for two and three distributed workers.  In another interesting line of work, Coded MapReduce has been proposed in \cite{LiMaAv:2015}, to reduce the communication between the mappers and reducers. However, the focus of this paper is significantly different, where we study the communication between the centralized master node and the distributed workers, motivated by the random reshuffling problem as initiated in \cite{Kannan-2015}.

\section{System Model}
\label{sec:Model}

We consider a master-worker distributed system, where a master node possesses the entire data-set. The master node sends batches of the data-set to the distributed workers over a shared link in order to locally calculate some function or train a model in a parallel manner. The local results are then fed-back to the master node, for iterative processing. In order to enhance the statistical performance of the learning algorithm, the data-set is randomly permuted at the master node before each epoch of the distributed algorithm, and then the shuffled data-points are transmitted to the workers.

We assume a master node which has access to the entire data-set $A=[x_1^T,x_2^T,\ldots,x_N^T]^T$ of size $Nd$ bits, i.e., $A$ is a matrix containing $N$ data points, denoted by $x_1,x_2,\ldots,x_N$, where $d$ is the dimensionality of each data point. Treating the data points $\{x_n\}$ as independent and identically distributed (i.i.d.) random variables, we have
\begin{subequations}
\begin{align}
H(x_n)=d, \;\;\forall n\in\{1,\ldots,N\}, \quad H(A)=Nd.\label{eq:data-set}
\end{align}
\end{subequations}

At each iteration, indexed by $t$, the master node divides the data-set $A$ among $K$ distributed workers, given as $A^{t}_{1}, A^{t}_{2}, \ldots, A^t_K$, where the batch $A^t_k$ is designated to be processed by worker $w_k$, and these batches correspond to the random permutation of the data-set, $\pi^t:A\rightarrow\{A^{t}_{1}, \ldots, A^t_K\}$.  Note that these data chunks are disjoint, and span the whole data-set, i.e.,
\begin{subequations}
\label{eq:data-batches}
\begin{align}
&A^t_i \cap A^t_j = \phi, \quad \forall i\neq j,\\
&A^t_1 \cup A^t_2 \cup \ldots \cup A^t_K =A, \quad \forall t.\label{eq:data-partitions}
\end{align}
\end{subequations}
Hence, the entropy of any batch $A^t_k$ is given as
\begin{align}
\label{eq:data-batches2}
H(A^t_k)= \frac{1}{K} H(A)= \frac{N}{K}d\quad ,\forall k\in\{1,\ldots,K\}.
\end{align}

After getting the data batch, each worker locally computes a function (as an example, this function could correspond to the gradient or sub-gradients of the data points assigned to the $k$th worker) $f_k(A^t_k)$, in iteration $t$,. The local functions from the $K$ workers are processed later at the master node, to get an estimate of the function $f_t(A)$. For processing purposes, the data block $A^t_k$ is needed to be stored by the worker while processing, therefore, we assume that worker $w_k$ has a cache $Z^t_k$ with storage capability of size $sd$ bits, for some real number $s$, that must at least store the data block $A^t_k$ at time $t$, i.e., if we consider $Z^t_k$ and $A^t_k$ as random variables then the storage constraint is given by
\begin{align}
\label{eq:cache-storage}
H(Z^t_k)=sd \geq H(A^t_k),\qquad \forall k\in\{1,\ldots,K\}.
\end{align}

For the scope of this paper, we focus on the setting of \textit{no-excess storage}, corresponding to $s= N/K$, in which each worker can exactly store $1/K$ fraction of the entire data, i.e., it only stores $s= N/K$ data points which are assigned to it in that iteration, therefore, the cache content at time $t$ for worker $w_k$ is given by $Z_k^t=A_k^t$, and the relationship in (\ref{eq:cache-storage}) is satisfied with equality. Henceforth, we drop the notation $Z^t_k$ as the cache content and use the notation for the data batch $A^t_k$ instead since they are the same for the no-excess storage setting. In the next epoch $t+1$, the data-set is randomly reshuffled at the master node according to the random permutation $\pi^{t+1}: A\rightarrow\{A^{t+1}_1, A^{t+1}_2, \ldots, A^{t+1}_K\}$.

The main communication bottleneck occurs during  {\textsf{\textit{Data Delivery}}} since the master node needs to communicate some function of the data to all the workers $X_{(\pi_t,\pi_{t+1})}$ of size $R_{(\pi_t,\pi_{t+1})}d$ bits, where $R_{(\pi_t,\pi_{t+1})}$ is the rate of the shared link based on the shuffle $(\pi_t,\pi_{t+1})$. Each worker $w_k$ should be able to extract the data points designated for it out of the incoming data, $X_{(\pi_t,\pi_{t+1})}$ from the master node as well as its locally stored data, i.e., $A^{t}_k$.

We next proceed to describe the data delivery mechanism, and the associated encoding and decoding functions. The main process then can be divided into 2 phases, namely the data delivery phase and the storage update phase as described next: in the \textit{data delivery phase}, the master node sends some function of the data to all the workers. Each worker should be able to extract the data points designated for it out of the incoming data from the master node as well as  the data stored in its local cache storage. In the \textit{cache update phase}, each worker stores the required data points for processing purposes, that can also be useful in reducing the communication overhead in subsequent epochs.


At time $t+1$, the master node sends a function of the data batches for the subsequent shuffles $(\pi_t,\pi_{t+1})$, $X_{(\pi_t,\pi_{t+1})} = \phi(A^{t}_1, \ldots, A^{t}_K,A^{t+1}_1,\ldots, A^{t+1}_K)=\phi_{(\pi_t,\pi_{t+1})}(A)$ over the shared link, where $\phi$ is the data delivery encoding function
\begin{equation}
\phi: \left[2^{\frac{N}{K}d}\right]^{2K} \rightarrow [2^{R_{(\pi_t,\pi_{t+1})}d}].
\end{equation}
Since $X_{(\pi_t,\pi_{t+1})}$ is a function of the data set $A$, we have
 \begin{subequations}
 \begin{align}
 &H\left(X_{(\pi_t,\pi_{t+1})}|A\right)=0,\label{eq:transmit-const}\\  
  &H\left(X_{(\pi_t,\pi_{t+1})}\right) =R_{(\pi_t,\pi_{t+1})}d.\label{eq:transmit-load}
 \end{align}
 \end{subequations}
Each worker $w_k$ should decode the desired batch $A^{t+1}_k$ out of the transmitted function $X_{(\pi_t,\pi_{t+1})}$, and the data stored in the previous time slot denoted as $A^{t}_k$. Therefore, the desired data is given by $A^{t+1}_k =\psi(X_{(\pi_t,\pi_{t+1})}, A^{t}_k)$, where $\psi$ is the decoding function at the workers
\begin{equation}
\psi: [2^{R_{(\pi_t,\pi_{t+1})}d}]\times [2^{sd}]\rightarrow [2^{\frac{N}{K}d}],
\end{equation} 
which also gives us the \textit{decodability constraint} as follows
\begin{equation}
\label{eq:decoding-const}
H\left(A^{t+1}_k|A^{t}_k, X_{(\pi_t,\pi_{t+1})}\right)=0 \quad ,\forall k\in\{1,\ldots,K\}.
\end{equation}
The update procedure for the no-excess storage setting is rather straightforward: worker $w_k$ keeps the part that does not change in the new shuffle, i.e., $A_k^{t+1}\cap A_k^{t}$. Then it removes the remaining part of its previously stored content, i.e., $A_k^{t}\setminus A_k^{t+1}$, and stores instead the new part, i.e, $A_k^{t+1}\setminus A_k^{t}$. 

Our goal in this work is to characterize the information theoretic bounds for optimal communication overhead $R_{(\pi_t,\pi_{t+1})}^*(K)$ for any arbitrary number of workers $K$, and any arbitrary shuffle $(\pi_t,\pi_{t+1})$, defined as
\begin{equation}
R_{(\pi_t,\pi_{t+1})}^*(K)=\underset{(\phi,\psi)}{\min}\quad \;R_{(\pi_t,\pi_{t+1})}^{(\phi,\psi)}(K),
\end{equation}
where $R^{(\phi,\psi)}_{(\pi_t,\pi_{t+1})}(K)$ is the rate of an achievable scheme defined by the encoding, and decoding functions $(\phi,\psi)$.
Subsequently, the optimal worst-case overhead is defined as
\begin{align}
&R_{\textsf{worst-case}} ^*(K)=\underset{(\pi_{t},\pi_{t+1})}{\max}\quad \;R_{(\pi_t,\pi_{t+1})}^*(K).
\end{align}

\section{Properties of Distributed Data Shuffling} 
Before presenting our main results on the communication overhead of shuffling, we present some fundamental properties that are satisfied for any two consecutive data shuffles give by $\pi_t: A \rightarrow \{A^t_1,\ldots ,A^t_K\}$, and $\pi_{t+1}: A\rightarrow \{A^{t+1}_1,\ldots ,A^{t+1}_K\}$. We start with the following definitions. 

\begin{definition}[Shuffle Index]\normalfont We define 
\begin{align}
\label{eq:S-defn}
S^{(\pi_t,\pi_{t+1})}_{i,j} \triangleq |A^{t}_{i}\cap A^{t+1}_j|,
\end{align}
as the shuffle index representing the number of data points that are needed by worker $w_j$ at time $t+1$, and are available at worker $w_i$ from the previous shuffle $t$.  
\end{definition}

\begin{definition}[Shuffle Matrix]\normalfont We also define the $K\times K$ shuffle matrix for the permutation pair $(\pi_t,\pi_{t+1})$ as 
\begin{align}
\label{eq:shuffle-matrix}
S^{(\pi_t,\pi_{t+1})}\triangleq \big[S^{(\pi_t,\pi_{t+1})}_{i,j}\big], \quad i,j\in\{1,\ldots,K\}.
\end{align}
\end{definition}
\vspace{5pt}

\begin{remark}\label{remark1} The significance of $S^{\pi_t,\pi_{t+1}}_{i,i}$ is that it is the number of common data points between $A^t_i$, and $A^{t+1}_{i}$. Thus, these number of data points do not need to be transmitted to worker $w_i$, and are not involved in the data delivery process.  Using the definition in (\ref{eq:S-defn}), together with (\ref{eq:data-batches}), it follows readily that 
\begin{align}
\label{eq:Sij-property}
&\overset{K}{\underset{i=1}{\sum}}S^{(\pi_t,\pi_{t+1})}_{i,j} = \sum_{i=1}^{K}|A^{t}_{i}\cap A^{t+1}_j| =    |A^{t+1}_{j}| = \frac{N}{K},\nonumber\\
&\overset{K}{\underset{j=1}{\sum}}S^{(\pi_t,\pi_{t+1})}_{i,j} = \sum_{j=1}^{K}|A^{t}_{i}\cap A^{t+1}_j| = |A^{t}_{i}| = \frac{N}{K}.
\end{align}
The properties in (\ref{eq:Sij-property}) imply that the sum of elements across any row (or column) for the shuffling matrix $S^{\pi_t,\pi_{t+1}}$ is constant for any shuffle $(\pi_t,\pi_{t+1})$ and is equal to $\frac{N}{K}$.
\end{remark}
\vspace{5pt}

\begin{remark}[Data-flow Conservation Property] \normalfont
We next state an important property satisfied by any shuffle, namely the data-flow conservation property:
\begin{equation}
\label{eq:dataflow-conservation}
\underset{j\in \{1,\ldots,K\}\setminus i}{\sum}S^{(\pi_t,\pi_{t+1})}_{j,i}= \underset{j\in \{1,\ldots,K\}\setminus i}{\sum}S^{(\pi_t,\pi_{t+1})}_{i,j}.
\end{equation}
The proof of this property follows directly from (\ref{eq:Sij-property}), and has the following interesting interpretation: the total number of new data points that need to be delivered to worker $w_i$ (and are present elsewhere), i.e., $\sum_{j\neq i}S^{(\pi_t,\pi_{t+1})}_{j, i}$ is exactly equal to the total number of data points that worker $w_i$ has that are desired by the other workers, which is  $\sum_{j\neq i}S^{(\pi_t,\pi_{t+1})}_{i, j}$. 
\end{remark}
\vspace{5pt}
\begin{definition}[Leftover Index and Leftover Matrix] \normalfont 
We define the leftover index as the number of leftover data-points needed by worker $w_j$ at time $t+1$ and available at $w_i$ at time $t$ as 
\begin{align}
\label{eq:leftover-defn}
\Omega^{\pi_t,\pi_{t+1}}_{i,j}\triangleq S^{\pi_t,\pi_{t+1}}_{i,j}-\min(S^{\pi_t,\pi_{t+1}}_{i,j},S^{\pi_t,\pi_{t+1}}_{j,i}).
\end{align}
The leftover matrix for the permutation pair $(\pi_t,\pi_{t+1})$ is defined as 
\begin{align}
\label{eq:leftover-matrix}
\Omega^{\pi_t,\pi_{t+1}}\triangleq [\Omega^{\pi_t,\pi_{t+1}}_{i,j}], \quad i,j\in\{1,\ldots,K\}.
\end{align}
\end{definition}
This definition and the significance of the leftover matrix will become clear in the subsequent sections, when we describe our proposed coded data delivery scheme. From the definition in (\ref{eq:leftover-defn}), we note that the diagonal entries of the leftover matrix are all zero.

\begin{remark}[Leftover Conservation Property] \normalfont
Analogous to the data-flow conservation property, we next show that the leftover indices also satisfy a similar leftover conservation property, as follows 
\begin{equation}
\label{eq:leftover-conservation}
\underset{j\in \{1,\ldots,K\}\setminus i}{\sum}\Omega^{(\pi_t,\pi_{t+1})}_{i,j}=\underset{j\in \{1,\ldots,K\}\setminus i}{\sum}\Omega^{(\pi_t,\pi_{t+1})}_{j,i}.
\end{equation}

To prove the above property, we use the definition of leftovers in (\ref{eq:leftover-defn}), to first compute the total leftovers at a worker $w_i$ as follows
\begin{align}
\label{eq:leftovers-total}
&\underset{j\in \{1,\ldots,K\}\setminus i}{\sum} \Omega^{(\pi_t,\pi_{t+1})}_{i,j}= \underset{j\in \{1,\ldots,K\}\setminus i}{\sum} S^{(\pi_t,\pi_{t+1})}_{i,j} \\
&\qquad\qquad\qquad-\underset{j\in \{1,\ldots,K\}\setminus i}{\sum} \min(S^{(\pi_t,\pi_{t+1})}_{i,j},S^{(\pi_t,\pi_{t+1})}_{j,i}).\nonumber
\end{align}
Similarly, we can also write the total number of leftover data points coming from all other workers to worker $w_i$
\begin{align}
\label{eq:needed-total}
&\underset{j\in \{1,\ldots,K\}\setminus i}{\sum} \Omega^{(\pi_t,\pi_{t+1})}_{j,i}= \underset{j\in \{1,\ldots,K\}\setminus i}{\sum} S^{(\pi_t,\pi_{t+1})}_{j,i}\\
&\qquad\qquad\qquad-\underset{j\in \{1,\ldots,K\}\setminus i}{\sum} \min(S^{(\pi_t,\pi_{t+1})}_{i,j},S^{(\pi_t,\pi_{t+1})}_{j,i}).\nonumber 
\end{align}
From the property in (\ref{eq:dataflow-conservation}), we notice that the quantities in (\ref{eq:leftovers-total}), and (\ref{eq:needed-total}) are equal and hence we arrive at the proof of (\ref{eq:leftover-conservation}). 
Using the leftover conservation property in (\ref{eq:leftover-conservation}), we can show that the sum across rows or columns for the leftover matrix $\Omega$ is constant for any shuffle $(\pi_t,\pi_{t+1})$.
\end{remark}

Subsequently, we refer to $R_{(\pi_t,\pi_{t+1})}$ as the rate for any achievable scheme $(\phi,\psi)$. We also drop the index $(\pi_t,\pi_{t+1})$ from $S^{(\pi_t,\pi_{t+1})}$, $\Omega^{(\pi_t,\pi_{t+1})}$, $R_{(\pi_t,\pi_{t+1})}$, and $X_{(\pi_t,\pi_{t+1})}$. 

\section{Main Results}
\label{sec:Results}
The  main contributions of this paper are presented next in the following three Theorems. 


\vspace{6pt}
\begin{theorem}
\label{th:1}
\textit{The optimal communication overhead $R^{*}(K)$ for a shuffle characterized by a shuffle matrix $S= [S_{i,j}]$ is upper bounded as}
\begin{align}
\label{eq:thm1}
 &R^*(K) \leq \overset{K-1}{\underset{i=1}{\sum}} \overset{K}{\underset{j=i+1}{\sum}}  \max(S_{i,j},S_{j,i})\nonumber \\
&\hspace{80pt} -\underset{k}{\max} \sum_{j\in \{1,\ldots, K\}\setminus \{k\}} \Omega_{k,j}.
\end{align}
\end{theorem}

\vspace{6pt}
\begin{theorem}
\label{th:2}
\textit{The optimal communication overhead $R^{*}(K)$, for any arbitrary shuffle matrix $S= [S_{i,j}]$ is lower bounded as}
\begin{equation}
\label{eq:thm2}
 R^*(K)\geq \overset{K-1}{\underset{i=1}{\sum}}\overset{K}{\underset{j=i+1}{\sum}}  S_{\sigma_i,\sigma_j},
\end{equation}
\normalfont
\textit{for any permutation $\sigma$: $\{1,\ldots,K\}\rightarrow \{\sigma_1,\ldots,\sigma_K\}$ of the $K$ workers.}
\end{theorem}

\vspace{6pt}
\begin{theorem}
\label{th:3}
\textit{The information theoretically optimal worst-case communication overhead for data shuffling is given by}
\begin{equation}
 R^*_{\textsf{worst-case}}(K) = \left(\frac{K-1}{K}\right)N.
\end{equation}
\end{theorem}

\vspace{5pt}

\section{Proof of Theorem \ref{th:1} \small{(Upper bound)}}
In this section, we present an achievable scheme for the shuffling process, which gives an upper bound on the communication overhead as stated in Theorem~\ref{th:1}.
We consider the random reshuffling process $(\pi_t,\pi_{t+1})$, characterized by a shuffle matrix $S=[S_{i,j}]$, from time $t$ given by the data batches $A^{t}_1,A^{t}_2,\ldots, A^{t}_K$, to time $t+1$ given by the data batches $A^{t+1}_1,A^{t+1}_2,\ldots, A^{t+1}_K$.
%

We first describe the main idea of our scheme through a representative example.

\vspace{5pt}
\hrule
\vspace{5pt}

\textit{\textbf{Example 1:}} Consider $K=3$ workers (denoted as $\{w_1, w_2, w_3\}$) and $N=15$ be the total number of data points. Consider the following shuffle matrix $S=[S_{i,j}]$:
\begin{equation}
\label{example:shuffle}
S=\left[\begin{array}{ccc}
S_{1,1} & S_{1,2} & S_{1,3}\\
S_{2,1} & S_{2,2} & S_{2,3}\\
S_{3,1} & S_{3,2} & S_{3,3}
\end{array} \right]= \left[\begin{array}{ccc}
2 & 1 & 2\\
2 & 1 & 2 \\
1 & 3 & 1
\end{array} \right] 
\end{equation}
The numbers in the diagonal represents the data points that remains unchanged across the workers, therefore, they do not participate in the communication process (see Remark~\ref{remark1}). 
For uncoded communication, the number of transmitted data points would be the sum of all non-diagonal entries, i.e., $R_{\textsf{uncoded}}=11.$

We first show how coding can be utilized to further reduce the communication overhead. For this example, worker $w_1$ needs $S_{2,1}=2$ data points  from $w_2$. Let us denote these points as  $\{x^{(1)}_{2,1}, x^{(2)}_{2,1}\}$.   At the same time, $w_2$ needs $S_{1,2}=1$ data point from $w_1$ (denoted as  $x_{1,2}$). Instead of uncoded transmission, the master node can send a coded symbol $x^{(1)}_{2,1}+ x_{1,2}$ which is simultaneously useful for both $w_1$, and $w_2$ as follows: $w_1$ has $x_{1,2}$, then it subtracts from the coded symbol to get the needed data-point $x^{(1)}_{2,1}$. Similarly, $w_2$ gets $x_{1,2}$ using $x^{(1)}_{2,1}$ and $x^{(1)}_{2,1}+ x_{1,2}$. 
This coded symbol is refereed to as an order-2 symbol, since it is useful for two workers at the same time.

By exploiting all such pairwise coding opportunities, we can send a total of $4$ order 2 symbols as follows: one coded symbol for $\{w_1,w_2\}$, one for $\{w_1,w_3\}$, and two for $\{w_2,w_3\}$. After having exhausted all pairwise coding opportunities, there are still some remaining data points, which we call as leftovers. The leftover matrix (defined in (\ref{eq:leftover-defn}) and (\ref{eq:leftover-matrix})), contains the number of leftover symbols after combining the order 2 symbols,  is given as
\begin{equation}
\label{example:leftover}
\Omega=\left[\begin{array}{ccc}
\Omega_{1,1} & \Omega_{1,2} & \Omega_{1,3}\\
\Omega_{2,1} & \Omega_{2,2} & \Omega_{2,3}\\
\Omega_{3,1} & \Omega_{3,2} & \Omega_{3,3}
\end{array} \right]= 
\left[\begin{array}{ccc}
0 & 0 & 1\\
1 & 0 & 0 \\
0 & 1 & 0
\end{array} \right]
\end{equation}
If the remaining $3$ leftover symbols (sum of all non-zero elements of $\Omega$) are sent uncoded, then, the total rate would be $R_{\textsf{paired-coding}}=R_{\textsf{coded-order2}}+R_{\textsf{uncoded-leftovers}}=4+3=7$, therefore, $R_{\textsf{paired-coding}}<R_{\textsf{uncoded}}$.

We now describe the main idea behind our proposed coding scheme which exploits a new type of coding opportunity as follows. Till this end, for each worker, we combine its incoming leftover symbols with its outgoing leftover symbols. By the leftover conservation property, these two are equal. Then, we have the three coded symbols as follows 
\begin{align}
\{x_{3,1}+x_{1,2}, \quad  x_{1,2}+x_{2,3}, \quad x_{2,3}+ x_{3,1} \}.
\end{align}
The key observation is that \textit{any two out of these three} coded symbols are enough for all the workers to get the remaining leftovers.
Two workers decode the needed points in one step, while the ignored worker decodes in two steps.

For example, if the master node transmits the first two coded symbols, i.e., $x_{3,1}+x_{1,2}$ and $x_{1,2}+x_{2,3}$, then the decoding works as follows:  $w_1$, and $w_2$ have $x_{1,2}$, and $x_{2,3}$, respectively, then they can get the needed ones, $x_{3,1}$, and $x_{1,2}$, respectively. Worker $w_3$, however, decodes its desired symbol through a two step procedure as follows: since it has $x_{3,1}$, then it can get $x_{1,2}$ from the first symbol $x_{3,1}+x_{1,2}$ in the first step. In the second step, from the second symbol $x_{1,2}+x_{2,3}$, it then uses $x_{1,2}$ to finally obtain the needed data point $x_{2,3}$. 
As a summary, we are able to send 3 leftovers in 2 coded symbols only. 
Therefore, communication overhead of the proposed scheme reduces to $R_{\textsf{proposed-coded}}=R_{\textsf{coded-order2}}+R_{\textsf{coded-leftovers}}=4+2=6$, i.e., $R_{\textsf{proposed-coded}}<R_{\textsf{paired-coding}}$. 

\vspace{5pt}
\hrule
\vspace{5pt}

We next present our proposed scheme for a general shuffle matrix and arbitrary number of workers $K$, which can be described in the following two phases, namely the first phase of transmitting order-2 symbols, and the second phase, which is what we call the leftover combining phase. 
\subsection{Phase 1: Order-2 symbols}
First we start by transmitting order-2 symbols, that are useful for two workers at the same time. If we consider two workers $w_i$, and $w_j$, then worker $w_i$ has some data points for worker $w_j$, given by $A_i^{t}\cap A_j^{t+1}$, which are $S_{i,j}=|A_i^{t}\cap A_j^{t+1}|$ data points in total. Similarly, $w_j$ has $S_{j,i}$ data points for $w_i$. Now, if we take all the data points $x_{i,j}\in A_i^{t}\cap A_j^{t+1}$, and combine them with the points $x_{j,i}\in A_j^{t}\cap A_i^{t+1}$ to transmit order-2 symbols jointly useful for $w_i$, and $w_j$, then we are limited by $\min(S_{i,j},S_{j,i})$ number of order-2 symbols for the pair $(i,j)$.
Therefore, we can transmit total number of order-2 symbols for all possible $(i,j)$ pairs of workers as follows
\begin{equation}
\label{eq:Phase1}
R_{\text{Phase 1}}=\overset{K-1}{\underset{i=1}{\sum}} \overset{K}{\underset{j=i+1}{\sum}} \min(S_{i,j},S_{j,i}).
\end{equation}

\subsection{Phase 2: Coded Leftover Communication} 
Now, we consider a coded approach for sending the leftovers after combining the order-2 symbols at phase 1. For a pair of workers $(i,j)$, after combining $\min(S_{i,j},S_{j,i})$ symbols in phase 1, then we still have $\Omega_{i,j}=S_{i,j}-\min(S_{i,j},S_{j,i})$ leftover symbols that are still needed to be transmitted from $w_i$ to $w_j$. Similarly, the leftovers form $w_j$ to $w_i$ is given by $\Omega_{j,i}=S_{j,i}-\min(S_{i,j},S_{j,i})$. We notice that if $S_{i,j}>S_{j,i}$, then $\Omega_{i,j}= S_{i,j}-S_{j,i}>0$, and $\Omega_{j,i}=0$, and vice versa. This gives us the following properties
\begin{align}
&\Omega_{i,j}+\Omega_{j,i}=\max(\Omega_{i,j},\Omega_{j,i})= |S_{i,j}-S_{j,i}|,\nonumber\\
&\min(\Omega_{i,j},\Omega_{j,i})=0.\label{eq:proberty-leftovers}
\end{align}
Clearly, if $S_{i,j}=S_{j,i}$, then $\Omega_{i,j}=\Omega_{j,i}=0$, and there are no leftover symbols for the pair $(i,j)$. The property in (\ref{eq:proberty-leftovers}) states that if a worker $w_i$ has some data points for $w_j$ in its leftovers ($\Omega_{i,j}\neq 0$), then $w_j$ has nothing in its leftovers needed by $w_i$ ($\Omega_{j,i}= 0$).
%
Using the leftover data conservation property in (\ref{eq:leftover-conservation}), we first state the following  claim:
\begin{claim}\label{claim2}
\textit{After combining the order-2 symbols in phase 1, the total number of symbols at a worker $w_i$ needed by other workers (outgoing leftovers) is equal to the total number of data points needed by the worker $w_i$ from other workers (incoming needed points).}
\end{claim}

As a simple scheme, we can use Claim~\ref{claim2} to combine all the leftovers with the needed data points for every worker $w_i$. Therefore, each worker can use its own outgoing leftover data points to get the desired incoming points. However, it is obvious that this coded scheme achieves the same rate as if we are sending the leftovers uncoded.

We next present the following claim which is one of the novel contributions of this paper:
\begin{claim}
\label{claim3}
\textit{If we combine the leftovers with the needed data points for any $K-1$ workers, then under a certain combining condition (stated below) for the remaining \textit{ignored worker}, say $w_k$, it can get its own needed data points without the need of being combined with its own leftovers.}
\end{claim} 

Before presenting the proof of Claim~\ref{claim3}, we first state the combining condition.
In order to ignore a worker $w_k$ from combining its leftovers with the needed points, the following condition must be satisfied while combining the leftovers with the needed points for other non-ignored workers:

\begin{definition}\textit{(Leftover Combining Condition for Ignoring $w_k$) \label{def:combining-condition}  The needed data-points at the ignored worker $w_k$ from leftovers of other workers $x_{i,k}$, and independently the leftovers at $w_k$ needed by other workers $x_{k,j}$ should only be combined with the data-points $x_{j,i}$ as follows}
\begin{align}
\left\{x_{k,j}+x_{j,i}, \:\: x_{j,i}+x_{i,k}\right\}.
\end{align}
\end{definition}

\begin{figure*}
\begin{center}
\includegraphics[width=0.95\textwidth]{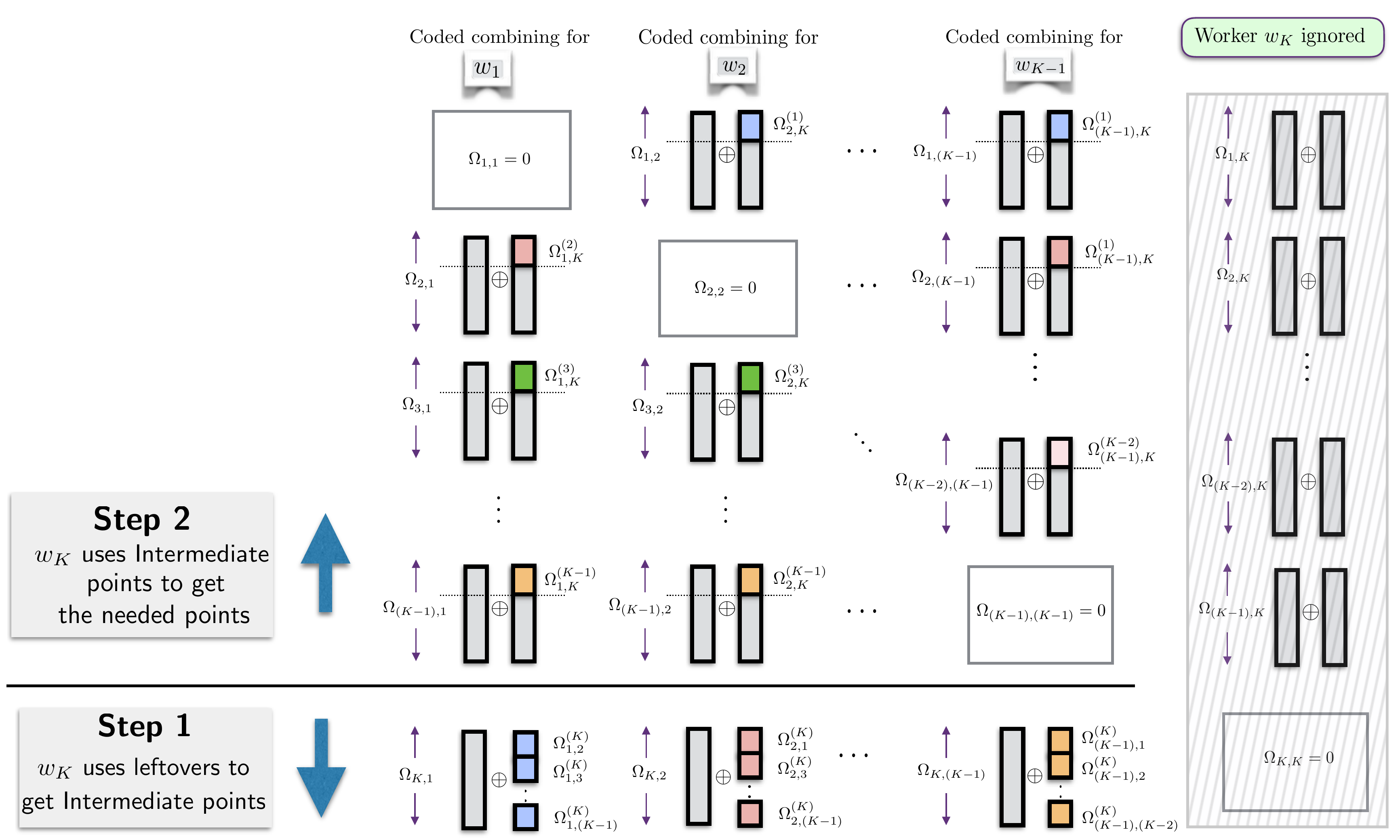}
\end{center}
\caption{\label{fig:leftover-combining} The leftover combining process after ignoring $w_K$. Below the solid line is the first step of decoding for the ignored worker where $w_K$ gets intermediate points using its leftover points. Above the solid line is the second step of decoding, where $w_K$ uses the intermediate points to decode the needed points.}
\end{figure*}

In order to understand the combining condition, we use the following example. Let us consider the following three types of leftover data points: (i) a data point $x_{i,k}$ that is needed by an ignored worker $w_k$, and is available at  worker $w_i$; (ii) a data point $x_{k,j}$ that is a leftover at $w_k$, and is needed by worker $w_j$; and (iii) a data point $x_{i,j}$ that is a leftover at $w_i$, and is needed by  worker $w_j$. 

In order for $w_k$ to decode $x_{i,k}$ using the leftover $x_{k,j}$, the leftover coded combining condition should be satisfied as follows

 $\bullet$  While combining the leftovers with the needed points of $w_j$ at the master node, the needed data point $x_{k,j}$ (from $w_j$'s perspective) should only be combined with the leftover data point $x_{j,i}$ as follows:
\begin{align}\label{eq:coded-sym1}
x_{k,j}+x_{j,i}.
\end{align}

 $\bullet$  While combining the leftovers with the needed points of $w_i$ at the master node, the leftover data point $x_{i,k}$ (from $w_i$'s perspective) should only be combined with the needed data point $x_{j,i}$ as follows:
\begin{align}\label{eq:coded-sym2}
x_{j,i}+x_{i,k}.
\end{align}

 From the above coded combining, we notice the following: 
1) Workers $w_i$, and $w_j$ still can decode the needed points  $x_{j,i}$, and $x_{k,j}$, respectively. 
2) Worker $w_k$ decodes in two steps: First, it uses $x_{k,j}$ to get $x_{j,i}$ from the coded symbol in (\ref{eq:coded-sym1}). In the next step, from the second coded symbol in (\ref{eq:coded-sym2}) it uses $x_{j,i}$ to decode the needed data point $x_{i,k}$.

\subsection{Proof of Claim~\ref{claim3}}
Now we need to prove formally the decodability at the ignored worker $w_k$. In order to complete the proof, we need to show that the number of intermediate points the ignored worker $w_k$ can get in the first step of decoding; are enough to decode the needed points in the next step of the decoding process.

We start by partitioning the leftover data points $\Omega_{i,j}$ into non-overlapping $(K-2)$ parts $\Omega_{i,j}^{(\ell)},\:\:\ell \in \{1,2,\ldots,K\}\setminus \{i, j\}$, where $\Omega^{(\ell)}_{i,j}\leq\Omega_{i,j}$ is defined as the number of intermediate (unintended since $\ell\neq\{i,j\}$) data points originally needed by $w_j$ that $w_\ell$ can get using its own leftovers needed for $w_i$ (through $w_i$). 

Therefore, $\Omega_{i,j}$ can be written as
\begin{align}
\label{eq:Omega-ij-struct}
\Omega_{i,j}=\underset{\ell\in\{1,\ldots,K\}\setminus\{i,j\}}{\sum}\Omega_{i,j}^{(\ell)}.
\end{align}
As shown in Figure~\ref{fig:leftover-combining}, $w_K$ for example uses its own leftovers needed by $w_1$ (through $w_1$), i.e., $\Omega_{K,1}$ points, to get unintended points (labelled with blue) that are needed by the other workers $\{2,3,\ldots,K-1\}$, i.e., $\Omega^{(K)}_{1,2},\ldots, \Omega^{(K)}_{1,K-1}$. Therefore, the total number of unintended (intermediate) data points recovered by $w_K$ using $\Omega_{K,1}$ data points is 
\begin{align}
\label{eq:Omega-ij-unintended}
\Omega_{K,1} = \overset{K-1}{\underset{j=2}{\sum}}\Omega^{(K)}_{1,j}.
\end{align}
Generally, through the combined symbols for $w_i$, the number of unintended data points which worker $w_\ell$ can obtain is
\begin{align}
\label{eq:dec-const1}
\Omega_{\ell,i} = \underset{j=\{1,\ldots,K\}\setminus\{i,\ell\}}{\sum}\Omega^{(\ell)}_{i,j}.
\end{align}

Let us assume now without loss of generality, that the ignored worker is the last worker $w_K$. As shown in Figure~\ref{fig:leftover-combining}, the ignored worker $w_K$ cannot get the needed data-points (colored chunks above the dotted lines) directly. Instead, $w_K$ uses its leftovers $\overset{K-1}{\underset{i=1}{\sum}}\Omega_{K,i}$ to get first unintended intermediate points (blue labelled points $\Omega^{(K)}_{1,j}$ through $w_1$, red labelled points $\Omega^{(K)}_{2,j}$ through $w_2$, etc.), which are shown below the solid line in the Figure.

In order for $w_K$ to make use of the intermediate symbols $\Omega^{(K)}_{i,j}$, $\{(i,j)\in\{1,\ldots ,K-1\},i\neq j\}$, every symbol $x_{i,j}$ of them should be paired up with data points useful for $w_K$ in the coded combining for $w_j$, i.e, $x_{i,j}+x_{j,K}$, which is satisfying the combining constraint in Definition~\ref{def:combining-condition}.
Following the relation in (\ref{eq:dec-const1}), the actual total number of unintended symbols $w_K$ can get in the first step of decoding is given by
\begin{align}
\label{eq:decK-intermediate}
\overset{K-1}{\underset{i=1}{\sum}}\Omega_{K,i}& = \overset{K-1}{\underset{i=1}{\sum}} \underset{j=\{1,\ldots,K-1\}\setminus\{i\}}{\sum}\Omega^{(K)}_{i,j}\nonumber\\
&=\underset{\substack{(i,j)\in\{1,\ldots,K-1\} \\i\neq j}}{\sum}\Omega^{(K)}_{i,j}.
\end{align}

 Using the unintended symbols that $w_K$ gets through $w_i$ and are originally needed by $w_j$, i.e., $\Omega_{i,j}^{(K)}$, it should be able to decode the needed symbols $\Omega^{(i)}_{j,K}$. As an example, $w_K$ gets the blue unintended data points  $\Omega^{(K)}_{1,2},\ldots, \Omega^{(K)}_{1,K-1}$ through $w_1$, then these data points are used to get the blue labelled needed points $\Omega^{(1)}_{2,K},\ldots, \Omega^{(1)}_{K-1,K}$ as shown above the solid line in Figure~\ref{fig:leftover-combining}.

The minimum number of unintended symbols $w_K$ needs to decode out of $\Omega_{i,j}$ points in the first step, should be enough to decode (equal to) the needed part $\Omega^{(i)}_{j,K}$ in the next step of decoding. From the unintended data recovery condition in (\ref{eq:dec-const1}), $\Omega^{(i)}_{j,K}$ is given by
\begin{align}
\label{eq:dec-const1-2}
\Omega^{(i)}_{j,K}=\Omega_{i,j} - \underset{\ell=\{1,\ldots,K-1\}\setminus\{i,j \}}{\sum}\Omega^{(i)}_{j,\ell}.
\end{align}
 
Therefore, the total number of unintended symbols that the worker $w_K$ should at least have in order to decode all the needed points in the next step is given by
\begin{align}
&\overset{K-1}{\underset{j=1}{\sum}}\Omega_{j,K}\overset{(a)}{=}
\underset{\substack{(i,j)\in\{1,\ldots,K-1\} \\i\neq j}}{\sum}\Omega^{(i)}_{j,K}\nonumber\\
&\overset{(b)}{=}\underset{\substack{(i,j)\in\{1,\ldots,K-1\} \\i\neq j}}{\sum}\Omega_{i,j} - \underset{\substack{(i,j,\ell)\in\{1,\ldots,K-1\} \\i\neq j \neq\ell}}{\sum}\Omega^{(i)}_{j,\ell}\nonumber\\
&\overset{(c)}{=}\underset{\substack{(i,j)\in\{1,\ldots,K-1\} \\i\neq j}}{\sum}\Omega_{i,j} - \underset{\substack{(i,j,\ell)\in\{1,\ldots,K-1\} \\i\neq j \neq\ell}}{\sum}\Omega^{(\ell)}_{i,j}\nonumber\\
&=\underset{\substack{(i,j)\in\{1,\ldots,K-1\} \\i\neq j}}{\sum}\left[\Omega_{i,j} - \underset{ \ell\in\{1,
\ldots,K-1\}\setminus\{i,j\}}{\sum}\Omega^{(\ell)}_{i,j}\right]\nonumber\\
&\overset{(d)}{=}\underset{\substack{(i,j)\in\{1,\ldots,K-1\} \\i\neq j}}{\sum}\Omega^{(K)}_{i,j}\label{eq:decK-intermediate-min},
\end{align}
where $(a)$ follows from (\ref{eq:Omega-ij-struct}), $(b)$ follows from the constraint in (\ref{eq:dec-const1-2}), $(c)$ by switching the sum indices, and $(d)$ from the definition in (\ref{eq:Omega-ij-struct}).
From (\ref{eq:decK-intermediate}) and (\ref{eq:decK-intermediate-min}), it now follows that the total number of intermediate points the ignored worker $w_K$ can decode in the first step is exactly equal to the minimum number it must decode in order to get the needed points in the second step, which completes the proof of Claim~\ref{claim3}.

Hence, the total communication overhead of phase $2$ is the total of all leftover symbols (except the ignored worker $k$), and is given as:
\begin{align}
&R_{\text{Phase 2}}=\overbrace{\underset{i\in\{1,\ldots,K\}\setminus \{k\}}{\sum}}^{\textsf{ignoring }w_k}\: \overbrace{\underset{j\in \{1,\ldots,K\}\setminus\{i\}}{\sum} \Omega_{i,j}}^{\textsf{leftovers at }w_i}\label{eq:variation}\\
&\:\:=\overset{K}{\underset{i=1}{\sum}}\: \underset{j\in \{1,\ldots,K\}\setminus\{i\}}{\sum} \Omega_{i,j}-\underset{j\in \{1,\ldots,K\}\setminus\{k\}}{\sum} \Omega_{k,j}\nonumber\\
&\:\:= \overset{K}{\underset{i=2}{\sum}}\: \overset{i-1}{\underset{j=1}{\sum}} \Omega_{i,j}+\overset{K-1}{\underset{i=1}{\sum}}\: \overset{K}{\underset{j=i+1}{\sum}} \Omega_{i,j}- \underset{j\in \{1,\ldots,K\}\setminus\{k\}}{\sum} \Omega_{k,j}\\
&\:\:\overset{(a)}{=} \overset{K-1}{\underset{i=1}{\sum}} \overset{K}{\underset{j=i+1}{\sum}} \Omega_{j,i}+\overset{K-1}{\underset{i=1}{\sum}}\: \overset{K}{\underset{j=i+1}{\sum}} \Omega_{i,j}- \underset{j\in \{1,\ldots,K\}\setminus\{k\}}{\sum} \Omega_{k,j}\\
&\:\:= \overset{K-1}{\underset{i=1}{\sum}} \overset{K}{\underset{j=i+1}{\sum}} \left(\Omega_{i,j}+\Omega_{j,i}\right)
- \underset{j\in \{1,\ldots,K\}\setminus\{k\}}{\sum} \Omega_{k,j}\\
&\:\:\overset{(b)}{=}\overset{K-1}{\underset{i=1}{\sum}} \overset{K}{\underset{j=i+1}{\sum}}\max(\Omega_{i,j},\Omega_{j,i})-\underset{j\in \{1,\ldots,K\}\setminus\{k\}}{\sum}\Omega_{k,j},\label{eq:Phase2}
\end{align}
where $(a)$ follows by swapping the indices $j$ and $i$ in the first summand, and $(b)$ follows from the property of leftovers in (\ref{eq:proberty-leftovers}), which states that that $\min(\Omega_{i,j},\Omega_{j,i})=0$.

Hence, the total communication overhead of the proposed scheme is the total number of transmitted symbols over Phases $1$ and $2$, which is the sum of (\ref{eq:Phase1}), and (\ref{eq:Phase2}), and is given by
\begin{align}
&R(K)= R_{\text{Phase 2}}+R_{\text{Phase 2}} \nonumber\\
&=\overset{K-1}{\underset{i=1}{\sum}} \overset{K}{\underset{j=i+1}{\sum}} \min(S_{i,j},S_{j,i})+\overset{K-1}{\underset{i=1}{\sum}} \overset{K}{\underset{j=i+1}{\sum}}\max(\Omega_{i,j},\Omega_{j,i})\nonumber\\
&\hspace{125pt} -\underset{j\in \{1,\ldots,K\}\setminus\{k\}}{\sum}\Omega_{k,j}\nonumber\\
&\overset{(a)}{=}\overset{K-1}{\underset{i=1}{\sum}} \overset{K}{\underset{j=i+1}{\sum}} \max(S_{i,j},S_{j,i})-\underset{j\in \{1,\ldots,K\}\setminus\{k\}}{\sum}\Omega_{k,j},\label{eq:Rate}
\end{align}
where $(a)$ follows from the property in (\ref{eq:proberty-leftovers}). 
In order to get the lowest possible rate for this scheme, which is also an upper bound for the optimal communication overhead, the choice of the ignored worker $w_k$ can be optimized to have the maximum number of leftovers, which is given by
\begin{align}
&R^*(K)\nonumber\\
&\leq \underset{k}{\min} \left( \:\:\overset{K-1}{\underset{i=1}{\sum}} \overset{K}{\underset{j=i+1}{\sum}} \max(S_{i,j},S_{j,i})-\underset{j\in \{1,\ldots,K\}\setminus\{k\}}{\sum}\Omega_{k,j}\right)\nonumber\\
&=\overset{K-1}{\underset{i=1}{\sum}} \overset{K}{\underset{j=i+1}{\sum}}\max(S_{i,j},S_{j,i})-\underset{k}{\max} \left(\underset{j\in \{1,\ldots,K\}\setminus\{k\}}{\sum}\Omega_{k,j}\right).
\label{eq:upper-bound}
\end{align}
This completes the proof of Theorem~\ref{th:1}.

\section{Proof of Theorem \ref{th:2} \small{(Lower bound)}}
In this section, we present the lower bound on the optimal communication overhead for any arbitrary random shuffle between two subsequent epochs $t$, and $t+1$ given by a shuffle matrix $S=[S_{i,j}]$, as stated in Theorem~\ref{th:2}.

\begin{align}
\label{eq:lower-bound}
Nd &\overset{(a)}{=}  H(A) \nonumber\\
&\overset{(b)}{=} I(A;A_1^{t},\ldots,A^{t}_K,X)+ H(A|A_1^{t},\ldots,A^{t}_K,X)\nonumber\\
&\overset{(c)}{=} H(A_1^{t},\ldots,A^{t}_K,X)-H(A_1^{t},\ldots,A^{t}_K,X|A)\nonumber\\
&\overset{(d)}{=} H(A_{\sigma_1}^{t},A^{t}_{\sigma_2},\ldots,A^{t}_{\sigma_K},X)\nonumber\\
&\overset{(e)}{=} H(A_{\sigma_K}^{t},X)+\overset{K-1}{\underset{i=1}{\sum}} H(A_{\sigma_i}^{t}|A_{\sigma_{i+1}}^{t},\ldots,A_{\sigma_K}^{t},X)\nonumber\\
&\overset{(f)}{\leq} H(A_{\sigma_K}^{t}) + H(X)+\overset{K-1}{\underset{i=1}{\sum}} H(A_{\sigma_i}^{t}|A_{\sigma_{i+1}}^{t+1},\ldots,A_{\sigma_K}^{t+1})\nonumber\\
&\overset{(g)}{\leq} \frac{Nd}{K}+Rd+\overset{K-1}{\underset{i=1}{\sum}} \left[\frac{Nd}{K}-I(A_{\sigma_i}^{t};A_{\sigma_{i+1}}^{t+1},\ldots,A_{\sigma_K}^{t+1})\right]\nonumber\\
&=Nd+Rd-\overset{K-1}{\underset{i=1}{\sum}}I(A_{\sigma_i}^{t};A_{\sigma_{i+1}}^{t+1},\ldots,A_{\sigma_K}^{t+1}),
\end{align}
where $(a)$ follows from (\ref{eq:data-batches2}), $(b)$ and $(c)$ are due to the fact that $I(A;B)=H(A)-H(A|B)=H(B)-H(B|A)$, and from (\ref{eq:data-partitions}) where the data-batches at any time span $A$, $(d)$ from (\ref{eq:data-partitions}) and (\ref{eq:transmit-const}), where the data-batches and $X$ are all functions of the data-set $A$, and $\sigma$ is any permutation of the the set $\{1,\ldots,K\}$, $(e)$ from the chain rule of entropy, $(f)$ from the decoding constraint in (\ref{eq:decoding-const}), the fact that conditioning reduces entropy, and the fact $H(A,B)\leq H(A)+H(B)$, and $(g)$ from (\ref{eq:data-batches2}), (\ref{eq:transmit-load}), and the fact $H(A|B)=H(A)-I(A;B)$. By rearranging the inequality in (\ref{eq:lower-bound}), we arrive at
\begin{align}
\label{eq:claim3}
Rd&\geq \overset{K-1}{\underset{i=1}{\sum}}I(A_{\sigma_i}^{t};A_{\sigma_{i+1}}^{t+1},\ldots,A_{\sigma_K}^{t+1})\nonumber\\
&= \overset{K-1}{\underset{i=1}{\sum}}\overset{K}{\underset{j=i+1}{\sum}}I(A_{\sigma_i}^{t};A_{\sigma_{j}}^{t+1})= \overset{K-1}{\underset{i=1}{\sum}}\overset{K}{\underset{j=i+1}{\sum}}S_{\sigma_i,\sigma_j} d.
\end{align}
Therefore, the lower bound on the communication overhead is given by $R^*(K)\geq \overset{K-1}{\underset{i=1}{\sum}}\overset{K}{\underset{j=i+1}{\sum}}S_{\sigma_i,\sigma_j}$, completing the proof of Theorem~\ref{th:2}.

\section{Proof of Theorem~\ref{th:3}}
In this section, we prove the optimality of our proposed scheme for the worst-case shuffle, which describes the maximum communication overhead across all possible shuffles.
\subsection{Achievability \small{(Worst-case Shuffle)}}
We start by using the upper bound described in Theorem~\ref{th:1}, where we use a variation of the expression in (\ref{eq:thm1}) by adding (\ref{eq:Phase1}), and (\ref{eq:variation}) as follows
\begin{align}\label{eq:upper-bound2}
R(K&)\overset{(a)}{=}\overset{K-1}{\underset{i=1}{\sum}} \overset{K}{\underset{j=i+1}{\sum}} \min(S_{i,j},S_{j,i})+\underset{i\in\{1,\ldots,K\}\setminus \{k\}}{\sum}\:\overset{K}{\underset{j=1}{\sum}} \Omega_{i,j}\nonumber\\
&\overset{(b)}{=}\overset{K-1}{\underset{i=1}{\sum}} \overset{K}{\underset{j=i+1}{\sum}} \min(S_{\sigma_i,\sigma_j},S_{\sigma_j,\sigma_i})+\overset{K-1}{\underset{i=1}{\sum}}\overset{K}{\underset{j=1}{\sum}} \Omega_{\sigma_i,\sigma_j}\nonumber\\
&\overset{(c)}{=}\overset{K-1}{\underset{i=1}{\sum}} \overset{K}{\underset{j=i+1}{\sum}} \min(S_{\sigma_i,\sigma_j},S_{\sigma_j,\sigma_i})+\overset{K-1}{\underset{i=1}{\sum}}\overset{K}{\underset{j=1}{\sum}} S_{\sigma_i,\sigma_j}\nonumber\\
&\quad-\overset{K-1}{\underset{i=1}{\sum}}\overset{K}{\underset{j=1}{\sum}}\min(S_{\sigma_i,\sigma_j},S_{\sigma_j,\sigma_i})\nonumber\\
&\overset{(d)}{\leq} \overset{K-1}{\underset{i=1}{\sum}}\overset{K}{\underset{j=1}{\sum}} S_{\sigma_i,\sigma_j}\overset{(e)}{=}\overset{K-1}{\underset{i=1}{\sum}} \frac{N}{K}=\left(\frac{K-1}{K}\right)N,
\end{align}
where $(a)$ holds because $\Omega_{i,i}=0$, $(b)$ follows by considering a permutation $\sigma=\{\sigma_1,\ldots\sigma_K\}$ of the workers, where $\sigma_K=k$ is the ignored worker,  $(c)$ follows from the definition of $\Omega_{i,j}$ in (\ref{eq:leftover-defn}), $(d)$ is due to the fact that $\min(S_{i,j},S_{j,i})\geq 0$, and $(e)$ from the property in (\ref{eq:Sij-property}).
Since this derived upper bound is found for any arbitrary shuffle, it is also an upper bound for the optimal worst-case communication overhead. Hence, we have
\begin{align}
\label{eq:wc-upperbound}
R^*_{\textsf{worst-case}}(K)\leq \left(\frac{K-1}{K}\right)N.
\end{align}

\subsection{Converse \small{(Information Theoretic lower bound)}}
We start by assuming a particular data shuffle, and then specialize our lower bound (obtained in Theorem~\ref{th:2}) for this particular shuffle. We use the fact that the worst-case overhead $R^{*}_{\textsf{worst-case}}(K)$ is lower bounded by the overhead of any shuffle $R(K)$, therefore the lower bound found for this given shuffle works as a lower bound for the worst-case as well, i.e.,
\begin{equation}
\label{eq:wc-to-any}
R^{*}_{\textsf{worst-case}}(K)\geq R^*(K).
\end{equation}
We assume a data shuffle matrix $S$ described as follows: For some permutation of the $K$ workers given by $\sigma=\{\sigma_1,\sigma_2,\ldots,\sigma_K\}$, any worker $w_{\sigma_{i+1}}$ at time $t+1$ needs only all the data points that $w_{\sigma_i}$ has from the previous shuffle at time $t$, which can be described as
\begin{equation}
\label{eq:wc-shuffle}
S_{\sigma_i,\sigma_j}=\left\{\begin{array}{cc}
\frac{N}{K}, & j=i+1,\\
0, & \text{otherwise}.
\end{array}\right.
\end{equation}
Therefore, using the lower bound in Theorem~\ref{th:2} given by (\ref{eq:thm2}), and using (\ref{eq:wc-to-any}), the lower bound for this particular shuffle, and hence the optimal worst-case shuffle, can be found as
\begin{align}
\label{eq:wc-lowerbound}
  R^*_{\textsf{worst-case}}(K)&\geq R^*(K)\geq  \overset{K-1}{\underset{i=1}{\sum}} \overset{K}{\underset{j=i+1}{\sum}} S_{\sigma_i,\sigma_j}\nonumber\\
   &=\overset{K-1}{\underset{i=1}{\sum}} S_{\sigma_i,\sigma_{i+1}}=\overset{K-1}{\underset{i=1}{\sum}} \frac{N}{K}=\left(\frac{K-1}{K}\right)N.
\end{align}
From (\ref{eq:wc-upperbound}), and (\ref{eq:wc-lowerbound}), it follows that the information theoretically optimal worst case communication overhead is 
\begin{equation}
\label{eq:wc-rate}
  R^*_{\textsf{worst-case}}(K)=\left(\frac{K-1}{K}\right)N.
\end{equation}

\section{Conclusion}
\label{sec:conclusion}
 In this paper, we presented new results on the minimum necessary communication overhead for the data shuffling problem.
We proposed a novel coded-shuffling scheme which exploits a new type of coding opportunity, namely coded leftover combining in order to reduce the communication overhead. Our scheme is applicable to any arbitrary shuffle, and for any number of distributed workers. We also presented an information theoretic lower bound on the optimal communication overhead that is also applicable for any arbitrary shuffle. Finally, we showed that the proposed scheme matches this lower bound for the worst-case communication overhead across all shuffles, and thus characterizes the information theoretically optimal worst-case overhead.

\bibliographystyle{IEEEtran}
\nocite{*}
\bibliography{allerton2016-arxiv.bbl}

\end{document}